# Nickel-based layered superconductor, LaNiOAs

Takumi Watanabe[a], Hiroshi Yanagi[a], Yoichi Kamihara[b], Toshio Kamiya[a,b], Masahiro Hirano[b,c], and Hideo Hosono[a,b,c,*]

[a] Materials and Structures Laboratory, Tokyo Institute of Technology, 4259 Nagatsuta, Midori-ku, Yokohama 226-8503, Japan

[b] ERATO-SORST, JST, in Frontier Research Center, Tokyo Institute of Technology, 4259 Nagatsuta, Midori-ku, Yokohama 226-8503, Japan

[c] Frontier Research Center, Tokyo Institute of Technology, 4259 Nagatsuta, Midori-ku, Yokohama 226-8503, Japan

**Abstract**

Rietveld analysis of the powder X-ray diffraction of a new layered oxyarsenide, LaNiOAs, which was synthesized by solid-state reactions, revealed that LaNiOAs belongs to the tetragonal ZrCuSiAs-type structure (*P*4/*nmm*) and is composed of alternating stacks of La-O and Ni-As layers. The electrical and magnetic measurements demonstrated that LaNiOAs exhibits a superconducting transition at 2.4 K, and above this, LaNiOAs shows metallic conduction and Pauli paramagnetism. The diamagnetic susceptibility measured at 1.8 K corresponded to ~20% of perfect diamagnetic susceptibility, substantiating that LaNiOAs is a bulk superconductor.

* Corresponding author. Fax: +81-45-924-5339, E-mail address:hosono@msl.titech.ac.jp

**1. Introduction**

The discovery of high transition temperature (high-$T_c$) Cu-based superconducting oxides has led numerous efforts to explore higher $T_c$ superconductors in a variety of materials, including Cu- and other transition metal-based oxides, because it is hypothesized that the strong electron correlation among 3d electrons plays an important role in Cu-based high-$T_c$ superconductors. Since then, many Cu-based high-$T_c$ superconductors have been discovered, including $HgBa_2Ca_2Cu_3O_{8+x}$ in 1993, which has $T_c$ of 133 K [1]. On the other hand, $T_c$'s of other transition metal based compounds (including oxides and pnictides) are lower than those of the Cu-based superconductors. However, the discoveries of superconductivity in new compounds such as $Sr_2RuO_4$ [2], $Na_xCoO_2 \cdot yH_2O$ [3], and a series of Ni-based layered borides [4,5,6] have provided complementary information to better understand the mechanism of superconductivity as well as clues to for exploring new material systems for higher $T_c$ superconductors.

Recently, we have investigated a series of quaternary compounds, La*MOPn* (*M*: transition metal element such as Mn, Fe, Co, and Ni, *Pn*: pnictogen element such as P and As) with the expectation that this materials system will be a new correlated electron system. In this crystal structure, the La-O and *M-Pn* layers are alternately stacked along the *c*-axis, as shown in Fig. 1. Systematic research on isostructural oxychalcogenides, LaCuO*Ch* (*Ch* = S, Se, Te), has indicated that the narrow energy gap Cu-*Ch* layer, which is sandwiched by larger energy gap La-O layers, works as a hole transport path [7]. Analogous to this, the *M-Pn* layer in La*MOPn* is also expected to be a carrier conduction path. This two-dimensional crystal structure with a transition metal element has led to the expectation that interesting electrical and/or magnetic properties, which originate from electron correlation, will be discovered.

In line with this strategy we found new superconductors, LaFeO*Pn* (*Pn* = P, As) [8,9] and LaNiOP [10], and an itinerant ferromagnet, LaCoO*Pn* (Pn = P, As) [11] in this materials system. Amongst them, LaFeOAs exhibited extremely interesting behaviors: the temperature dependence of the resistivity of undoped LaFeOAs showed a semiconductor-like behavior with a kink in resistivity at ~150 K and a superconducting transition did not occur, but its electrical property changed to a metallic behavior by F (electron) doping. Moreover, LaFeOAs with a F content greater than or equal to ~4 at.% showed superconducting transitions. Its transition temperature reached ~26 K around a F content of 11 at.%. On the other hand, LaFeOP and LaNiOP appeared to be conventional superconductors with $T_c$'s lower than 5 K. Accordingly the superconducting properties of LaFeO*Pn* were drastically altered by changing the *Pn* element from P to As. Thus, elucidating the electrical and magnetic properties of LaNiOAs is important for systematic and comparative studies of the LaFeO*Pn* and LaNiO*Pn* systems. However, a study on the synthesis of LaNiOAs has yet to be reported.

In this letter, we report the synthesis as well as the electrical and magnetic properties of LaNiOAs. LaNiOAs exhibits metallic conduction at high temperatures and a superconducting transition at 2.4 K without impurity doping, which contrasts the LaFeO*Pn* case, suggesting that the semiconductor-like behavior in the undoped sample and the induction of the superconducting transition by F doping are characteristic of the LaFeO*Pn* system. The present results along with those reported for the La*M*O*Pn* system would help elucidate the mechanism of superconductivity in this system.

2. Experimental

Polycrystalline samples of LaNiOAs were prepared by solid-state reactions of the starting materials, La (Shin-etsu Chemical, purity 99.5%), As (Kojundo Chemical Laboratory, 99.99%), and NiO (Kojundo Chemical Laboratory, 99.97%). Single-phase LaAs was initially synthesized by heating a stoichiometric mixture of La metal powder and As in an evacuated silica tube at 400 °C for 12 h and 700 °C for 6 h. Then a stoichiometric mixture of LaAs and NiO was pressed into a pellet and heated at 1000 °C for 1 day in an evacuated silica tube. The resulting sample was characterized by high-power X-ray diffraction (XRD, D8 ADVANCE-TXS, Bruker AXS) with Cu K$\alpha$ radiation at 24 °C. Multi-phase powder X-ray Rietveld analyses were carried out using the code TOPAS3 [12] to refine the crystallographic parameters of LaNiOAs and to estimate weight fractions of LaNiOAs and the impurity phases. The initial parameters were taken from an isostructural compound, LaNiOP [10].

The electrical resistivity of the sintered pellets was measured in the temperature range of 1.8 to 305 K by a four-probe technique (using PPMS, Quantum Design). Ohmic contact electrodes were formed by Au sputtering and electrical connections were made using a silver paste. The magnetic properties were measured with a vibrating sample magnetometer (VSM, using PPMS, Quantum Design). The temperature dependence of the magnetic susceptibility was measured at temperatures from 1.8 to 305 K in a magnetic field of 10 Oe after zero field cooling. The field dependence of magnetization was measured at 1.8 K.

3. Result and discussion
3.1 XRD and Rietveld analyses

The obtained samples were chemically stable in air and dark gray. Fig. 2 shows the XRD patterns measured (black circles) and simulated by Rietveld analysis (red line) for the purest sample. Rietveld analyses revealed that the samples contained small amounts of impurity phases, including NiAs, $La_2O_3$, and $La(OH)_3$. $La(OH)_3$ was probably produced by hydration of $La_2O_3$ after the samples were taken from the silica tube to the atmosphere. Four-phase Rietveld analysis provided a weight fraction of LaNiOAs in the purest sample of 99 wt%. Table 1 summarizes the refined crystallographic parameters of LaNiOAs. Rietveld analysis also substantiated that LaNiOAs has the same crystal structure as La*M*O*Pn* (*M* = Mn, Fe, Co, Ni, Zn, *Pn* = P, As), which belongs to the tetragonal ZrCuSiAs-type structure (*P*4/*nmm*). The crystal structure is composed of alternating stacks of Ni-As and La-O layers, which are built from the edge-sharing networks of $NiAs_4$ and $OLa_4$ tetrahedrons, respectively. These tetrahedrons are distorted and more planar compared to regular tetrahedron, and are characterized by the two nearest-neighbor bonding angles. The As-Ni-As bond angles are 122.95° and 103.18°, and those of La-O-La are 119.45° and 104.73°. The large deviation of the bond angles from that of regular tetrahedron (109.47°) have been observed commonly in Ni- and Fe-based ZrCuSiAs-type compounds: 126.4° and 101.7° for LaNiOP [10], and 120.2° and 104.4° for LaFeOP [8]. While, those of e.g. LaCuO$Ch$ ($Ch$ = S and Se), which are isostructural compounds but have not exhibited a superconducting transition [13], are much closer to that of regular tetrahedron. It implies that the distortion of the $MPn_4$ tetrahedron has a close relation to carrier transport including a superconducting transition, but further investigation is required to find a systematic relationship.

### 3.2 Electrical and magnetic properties

Fig. 3 shows the temperature ($T$) dependence of the electrical resistivity ($\rho$). $\rho$ exhibited a metallic behavior at temperatures down to 2.4 K, and the $\rho$ value at 305 K was very low, 3.5 mΩ•cm. The inset shows a magnified view of $\rho$-$T$ curves as a function of the external magnetic field ($H_{ext}$) in the temperature range of 1.8 to 2.8 K. The resistivity began to decrease at 2.4 K and became zero around 2 K. The onset temperature shifted to a lower temperature as $H_{ext}$ increased, and the drop in $\rho$ vanished at $H_{ext}$ = 3000 Oe. These results suggest that LaNiOAs exhibits a superconducting transition at 2.4 K.

Fig. 4 shows that the temperature dependence of the mass magnetic susceptibility ($\chi_g$) measured at 10 Oe after zero-field cooling. $\chi_g$ was $0.99 \times 10^{-6}$ emu/g at 305 K, and was nearly independent of temperature between 2.4 and 305 K, implying Pauli paramagnetism. However, $\chi_g$ began to decrease and became negative at 2.4 K. It reached a large negative value of $-1.61 \times 10^{-3}$ emu/g at 1.8 K. These results, along with the zero resistance, clearly indicate that LaNiOAs exhibits superconductivity at temperatures below 2.4 K. The transition temperature is decreased by replacing $Pn$ from P ($T_c$ = ~3 K) to As (~2.4 K), while the $\rho$ value at 305 K is increased from 1.3 to 3.5 mΩ•cm. This result on the transition temperature also makes a sharp contrast to the LaFeO$Pn$ system: LaFeOP shows a superconducting transition at ~5 K but replacement of P with As raises it to ~26 K when F is doped (i.e. upon electron doping). The absolute value of the $H_{ext}$ dependence of magnetization ($M$), shown in the inset, proportionally increased with $H_{ext}$ at $H_{ext}$ less than 15 Oe, but decreased to zero as $H_{ext}$ approached 400 Oe. This behavior is similar to that observed in type-II superconductors. In this case, the lower ($H_{c1}$) and upper critical fields ($H_{c2}$) were estimated to be ~15 Oe and ~400 Oe, respectively. The $\chi_g$ value estimated from the proportional region in the $M$-$H_{ext}$ curve corresponded to ~20% of perfect diamagnetism, which indicates that almost 80% of the sample is not in the superconducting state at the lowest measurement temperature 1.8 K.

Hence, the total amount of the impurity phases by weight fraction estimated from the Rietveld analysis was ~1 wt.%, which is negligible compared to that of the volume fraction of the superconducting phase (~20% at 1.8 K). In addition, among the impurity phases detected in the XRD pattern, only NiAs shows an electronic conduction, but Matthias and Hulm have reported that NiAs does not show a superconducting transition down to 1.3 K [14]. Accordingly, we conclude that LaNiOAs is a superconductor with $T_c$ of 2.4 K.

## 4. Conclusions

Solid state reactions were used to synthesize a new layered oxyarsenide, LaNiOAs. Powder X-ray Rietveld analysis revealed that LaNiOAs has a tetragonal ZrCuSiAs-type structure and is composed of alternating stacks of La-O and Ni-As layers. LaNiOAs exhibited a superconducting transition at $T_c$ = 2.4 K. The present study adds a new superconductor LaNiOAs to the La*M*O*Pn* system, which already includes superconductors such as LaFeOP, LaNiOP, and F-doped LaFeOAs, and the addition of LaNiOAs leads to the systematic conclusion that only Fe- and Ni-based La*M*O*Pn* exhibit superconducting transitions at temperatures down to 1.8 K among these types of compounds synthesized to date (*M* = Mn, Fe, Co, Ni, Zn, *Pn* = P, As). On the other hand, the arsenides show different behaviors for LaFeOAs and LaNiOAs. LaFeOAs shows a semiconductor-like behavior and a superconducting transition is induced when F is doped (electron doping), whereas LaNiOAs shows a metallic behavior and a superconducting transition occurs in undoped samples. Hence, we think that an extended systematic study on La*M*O*Pn*, where the La, *M*, and *Pn* are replaced with other lanthanides, transition metals, and N group elements, respectively, is important to discover new superconductors as well as to elucidate the mechanism of superconductivity.

We like to comment that many preprint papers on this materials system have been reported while this letter was reviewed: higher-$T_c$ superconductors were realized by replacing La by rare earth elements having incomplete 4f shells [15]. These results support our above expectation.

Table 1. Crystallographic parameters refined by powder X-ray Rietveld analysis for the LaNiOAs sample. X-ray diffraction data was collected at 24 °C. Occupancy and the $B$ values are fixed to 1.0 and the values in the table, respectively. Reliability factor is $R_{wp}$ = 7.52%, and the goodness of fit parameter is $S$ = 2.02. The bond angles of $\alpha$ and $\beta$ are illustrated in the inset of Figure 2.

| $a$ (nm) | $c$ (nm) | $V$ (nm$^3$) |
|---|---|---|
| 0.412309(1) | 0.818848(6) | 0.139203(1) |

| Atom | Wycoff position | $x$ | $y$ | $z$ | $B$ (Å$^2$) |
|---|---|---|---|---|---|
| La | 2c | 1/4 | 1/4 | 0.14697(9) | 0.25 |
| Ni | 2b | 3/4 | 1/4 | 1/2 | 0.40 |
| O | 2a | 3/4 | 1/4 | 0 | 0.90 |
| P | 2c | 1/4 | 1/4 | 0.6368(1) | 0.45 |

| | $d$ (nm) | | $\alpha$ (°) | $\beta$ (°) |
|---|---|---|---|---|
| La-O | 2.3871(4) | La–O–La | 104.73(2) | 119.45(3) |
| Ni-As | 2.3463(7) | As–Ni–As | 103.18(3) | 122.95(6) |

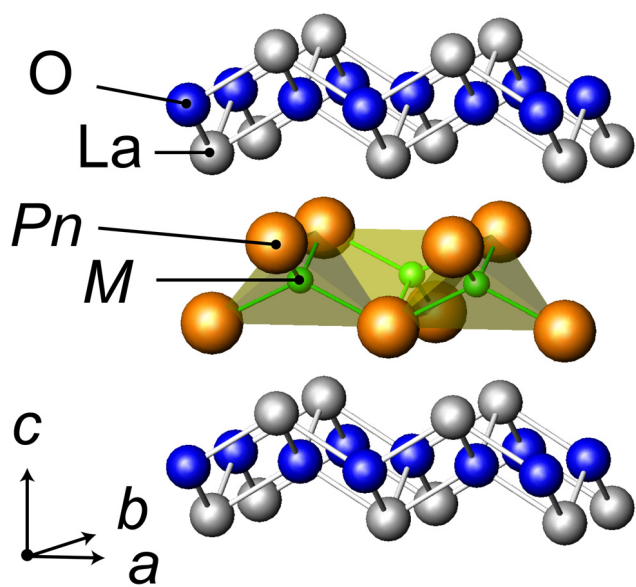

Fig. 1. Crystal structure of La*M*O*Pn* (*M* = Mn, Fe, Co, Ni, Zn, *Pn* = P, As). La-O and *M*-*Pn* layers are alternately stacked along the *c*-axis.

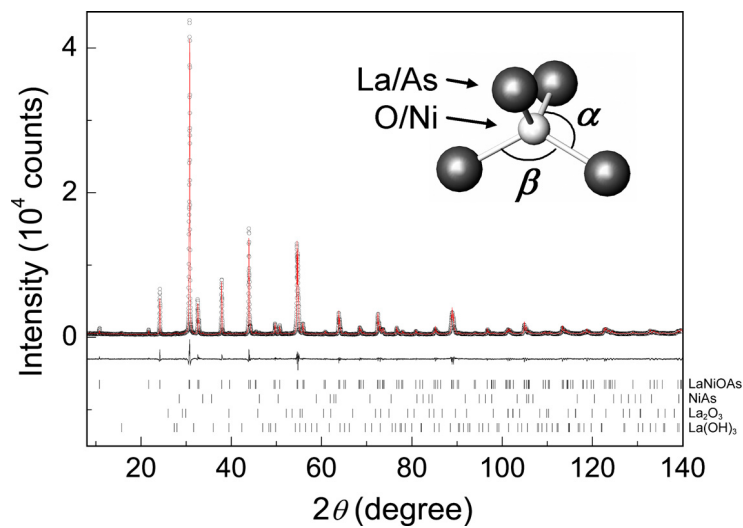

Fig. 2. XRD patterns of LaNiOAs measured (black circles) and simulated by Rietveld analysis (red lines). Line and the vertical bars below the XRD patterns show the difference profile between the observed and simulated XRD patterns, and the calculated positions of the Bragg diffractions of LaNiOAs, NiAs, $La_2O_3$, and $La(OH)_3$ from top to bottom, respectively. The inset shows a structure of the $NiAs_4$ and $OLa_4$ tetrahedrons to identify the bond angles of $\alpha$ and $\beta$ that are summarized in Table 1.

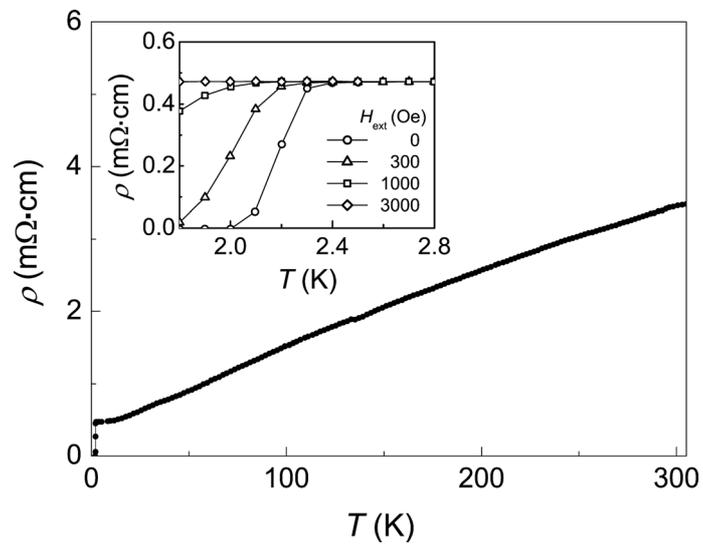

Fig. 3. Temperature (*T*) dependence of the electrical resistivity ($\rho$) at 0 Oe. Inset shows the $\rho$-*T* curves as a function of the external magnetic field ($H_{ext}$) magnified in the temperature range of 1.8 to 2.8 K.

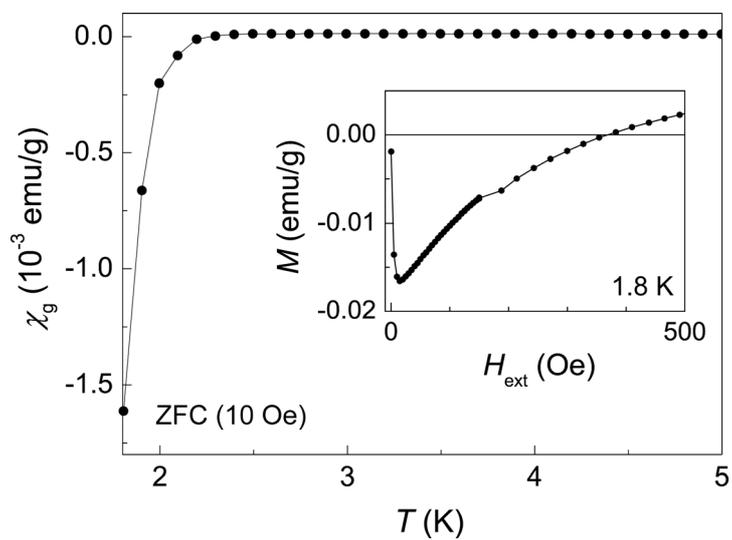

Fig. 4. Temperature (*T*) dependence of magnetic susceptibility ($\chi_g$) in the temperature range of 1.8 to 5 K measured after zero-field cooling at 10 Oe. Inset shows the external magnetic field ($H_{ext}$) dependence of magnetization (*M*) at 1.8 K.